\documentclass[pra,a4paper, aps,showpacs,superscriptaddress,groupedaddress]{revtex4}
\usepackage{ae}
\usepackage[T1]{fontenc}
\usepackage[ansinew]{inputenc}
\usepackage{amsmath}
\usepackage{amssymb}
\usepackage[caption=false]{subfig}
\usepackage{multirow}
\usepackage{array}
\usepackage[]{graphicx}
\usepackage{wrapfig}

\usepackage{color}
\usepackage[colorlinks]{hyperref}
\usepackage{lscape}
\begin{document}
\title{Effect of local filtering on Freezing Phenomena of Quantum Correlation}
\author{Sumana Karmakar}
\email{sumanakarmakar88@gmail.com}
\affiliation{Department of Applied Mathematics, University of Calcutta, 92, A.P.C. Road, Kolkata-700009, India.}
\author{Ajoy Sen}
\email{ajoy.sn@gmail.com}
\affiliation{Department of Applied Mathematics, University of Calcutta, 92, A.P.C. Road, Kolkata-700009, India.}
\author{Amit Bhar}
\email{bhar.amit@yahoo.com}
\affiliation{Department of Mathematics, Jogesh Chandra Chaudhuri College, 30, Prince Anwar Shah Road, Kolkata-700033, India.}
\author{Debasis Sarkar}
\email{dsappmath@caluniv.ac.in,debasis1x@yahoo.co.in}
\affiliation{Department of Applied Mathematics, University of Calcutta, 92, A.P.C. Road, Kolkata-700009, India.}

\begin{abstract}
General quantum correlations measures like quantum discord, one norm geometric quantum discord, exhibit freezing, sudden change, double sudden change behavior in their decay rates under different noisy channels. Therefore, one may attempt to investigate how the freezing behavior and other dynamical features are affected under application of local quantum operations. In this work, we demonstrate the effect of local filtering on the dynamical evolution of quantum correlations. We have found that using local filtering one may remove freezing depending upon the filtering parameter. 
\end{abstract}
\date{\today}
\maketitle
\section{Introduction}
Quantification and characterization of quantum correlations that can not be fully captured through entanglement measures has generated  lot of interests in recent days. It has been shown that a completely separable mixed state might show a  quantum signature to compute information processing tasks \cite{seperable}. Among several general non-classical correlation measures, Quantum Discord (in short, QD) \cite{QD} draws much attention due to its operational significance in various  quantum information processing tasks, like quantum metrology \cite{metrology1,metrology2,metrology3}, entanglement activation \cite{entanglement activation1,entanglement activation2,entanglement activation3}, information encoding and distribution \cite{encoding1,encoding2}, etc. In order to reveal quantumness in several composite quantum systems beyond entanglement, many attempts are made to show differences between discord like measures with entanglement. One of these attempts to study dynamics of general quantum correlations in open quantum systems.

Several peculiar properties in the dynamics of classical and quantum correlations have  already been established in the presence of Markovian and Non-Markovian noise\cite{Markovian noise 1,Markovian noise 2,Markovian noise 3,Markovian noise 4,Markovian noise 5,Markovian noise 6,Markovian noise 7,Non markovian noise}. It has been  shown  that under dissipative dynamics where entanglement suddenly disappears, known as entanglement sudden death \cite{ESD}, quantum discord vanishes only in the asymptotic limit. In this sense, quantum discord  is more robust against decoherence  than entanglement \cite{robustness1,robustness2}. Under Markovian noise quantum discord exhibits some striking phenomena in its decay rate for  Bell diagonal states such as freezing \cite{freezing of QD,freezing of QC}, single sudden change \cite{sudden change of QD}. This freezing phenomena, not exhibited by any entanglement measure, is very demanding since it indicates that the quantum protocols in which quantum correlations are used as resources, will run with a performance unaffected by specific noisy conditions. Thus more intensive study of the behavior of quantum discord under different noisy channel is highly important.

Alternative to the entropic approach, quantum correlation can be measured in geometric way. Recently, Cianciaruso \textit{et.al.}\cite{universality of freezing} have proved that this freezing phenomena occurs for any geometric measure of quantum correlation whenever the distance defining the measure respects a minimal set of physical assumptions, namely dynamical contractivity under quantum channels, invariance under transposition, and convexity. Thus freezing phenomena is revealed as universal property for such geometric measures of quantumness. The examples of such distances are the relative entropy\cite{relative entropy1,relative entropy2}, the squared bures distance\cite{burace distance1,burace distance2,freezing of QC,burace distance3}, the squared Hellinger distance\cite{hellinger1,hellinger2,hellinger3} and the trace (or Schatten one-norm) distance\cite{trace distance1,1-GQD,trace distance2}. The Hilbert-Schmidt distance(or Schatten two-norm)\cite{GQD1} does not respect the contractivity property\cite{GQD problem1} and as a result the geometric quantum discord(GQD)\cite{GQD2}, based on Hilbert-Schmidt distance may increase under local reversible operation on unmeasured party\cite{GQD problem2}. Thus in-spite of its computational simplicity and operational significance \cite{operational significance of GQD} in quantum communication protocol, GQD is not considered as a good measure of non classicality. Another version of geometric discord, based on Schatten one-norm, was introduced by Sarandy \textit{et.al.}\cite{1-GQD}. This measure is more acceptable to us since it does not suffer from the problems like its Schatten two counterpart and  also for its computational simplicity. This measure displays freezing\cite{freezing of QC}, single sudden change\cite{sudden change of 1-GQD} behaviors for Bell diagonal states under decoherence like quantum discord. It has been shown that under Markovian noise, one-GQD exhibits twice transition in its decay rate namely double sudden change\cite{sudden change of 1-GQD} for Bell diagonal states. Now, one may ask whether it is possible to maintain freezing (or other effects) or remove it under some local quantum operations or not. In this work, we have addressed this issue under local filtering operations.

Behavior of local filtering operations on entanglement has been studied earlier. A filter can be used for creation as well as purification of entanglement\cite{purification1,purification2}. It is possible to retrieve entanglement by a single local filtering for initially pure W and Cluster states with general amplitude damping channel as noise model\cite{purification3}. Here, we will explore the effects of single local filtering operation on dynamical evolution of quantum discord and one-GQD in noisy environment. We will analyze the effect of single local filtering against freezing, sudden change and double sudden change. We consider standard Bell diagonal state as initial state and \textit{phase flip} (PF), \textit{bit flip} (BF), \textit{bit-phase flip} (BPF) channels as our noise models for decoherence.

Our paper is organized as follows: in section II we will discuss some measures of quantum correlations beyond entanglement which we will use in our work. In section III we will discuss about behavior of quantum correlations for Bell diagonal states under decoherence and section IV contains our main results. Finally, we conclude in section V.

\section{Measures of Quantum Correlations}

\paragraph{\textbf{Quantum Discord:}}
Olliver and Zurek \cite{QD} introduced the concept of \textit{Quantum Discord} as a measure of genuine quantum correlation. The total correlation, i.e., the total amount of classical and quantum correlations of a bipartite state $\rho_{AB}$ is given by its quantum mutual information
\begin{equation}\label{mutual information}
I(\rho_{AB})=S(\rho_A)+S(\rho_B)-S(\rho_{AB})
\end{equation}
where $S(\rho)=-Tr(\rho \log_2\rho)$ is the Von Neumann entropy of a state $\rho$
and $\rho_A$, $\rho_B$ are the reduced density matrices of the state $\rho_{AB}$.
On the other hand, its classical correlation is captured by \cite{CC}
\begin{equation}\label{classical correlation}
C(\rho_{AB})=\max_{\{\Pi^j_A\}}[S(\rho_B)-S(\rho_{AB}|\{\Pi^j_A\})]=\max_{\{\Pi^j_A\}}[S(\rho_B)-\sum_j {p_j S(\rho_{B/j})}]
\end{equation}
where maximum is taken over all complete set of orthogonal projectors $\{\Pi^j_A\}$ on subsystem $A$ and $\rho_{B/j}=tr_A[\Pi^j_A\otimes I \rho_{AB}\Pi^j_A\otimes I ]$ is the reduced density matrix of the subsystem $B$ after obtaining measurement outcome $j$ with probability $p_j=Tr_{AB}[\Pi^j_A\otimes I \rho_{AB}\Pi^j_A\otimes I]$. Then \textit{Quantum Discord} (QD) of a bipartite state $\rho_{AB}$ is defined as the difference between its total correlation and classical correlation and is given by
\begin{equation}\label{QD}
 Q(\rho_{AB})=I(\rho_{AB})-C(\rho_{AB})
\end{equation}
This is possibly the most important quantifier of quantum correlations beyond entanglement. There are separable states with non-zero quantum discord. The zero discord states have the form $\rho_{AB}= \sum_i p_i |i\rangle\langle i|^A\otimes \rho_i^B$, where $p_i$ is a probability distribution, $\{|i\rangle^A \}$ denotes an orthonormal basis for subsystem $ A$ and $\rho_i^B$ is an ensemble of states of subsystem $B$. Zero-discord states are usually known as classical-quantum states and the set of all zero-discord state is not convex unlike set of all separable states. For this reason in general it is really hard to calculate quantum discord for most of the states. Only few results are available. For pure bipartite states quantum discord coincides with entropy of entanglement.
\paragraph{\textbf{One-norm Geometric Quantum Disord :}}
Let us consider the geometric quantum discord based on more general norm
\begin{equation}\label{p-GQD}
 D_p = \min_{\Omega_0}{\|\rho_{AB}-\rho_{AB}^c\|_p^p},
\end{equation}
where $\|X\|_p=Tr[(X^\dagger X)^{\frac{p}{2}}]^{\frac{1}{p}}$ is the Schatten $p$-norm with $p$ as positive integer and $\Omega_0$ is the set of classical quantum states. In this notation the geometric quantum discord (GQD) introduced by Dakic \textit{et.al.}\cite{GQD2} is simply obtained by taking $p=2$. In-spite of its computation simplicity it fails to establish as a good quantifier of quantum correlation since it may increase under local reversible operation on unmeasured party \cite{GQD problem2}. Sarandy \textit{et.al.} \cite{1-GQD} have shown that only One-norm Geometric Quantum Discord (1-GQD) is the only possible Schatten $p$-norm which does not suffer from this local ancillary problem. The one-norm Geometric Quantum Discord of a bipartite state $\rho_{AB}$ is defined as \cite{1-GQD}
\begin{equation}\label{1-GQD}
 D_G=D_1 = \min_{\Omega_0}{\|\rho_{AB}-\rho_{AB}^c\|_1},
\end{equation}
where $\|X\|_1=Tr[\sqrt{X^\dagger X}]$ is the trace norm.\\ 

Like Quantum Discord, 1-GQD is zero if and only if $\rho_{AB}$ is classical-quantum state. It
is invariant under local unitary operations. For pure bipartite states One-norm Geometric Quantum Discord is an entanglement monotone.

\section{Dynamics of quantum correlation for Bell Diagonal states under decoherence:}

Any standard Bell diagonal state can be written as
\begin{equation}\label{bell diagonal state}
\rho=\frac{1}{4}[I\otimes I+\sum_{i=1}^3{c_i\sigma^i\otimes \sigma^i}]=\sum_{i=1}^4{\lambda_i|\phi_i\rangle\langle\phi_i|}
\end{equation}
where $|\phi_{1,3}\rangle=\frac{1}{\sqrt2}(|00\rangle\pm|11\rangle)$, $|\phi_{2,4}\rangle=\frac{1}{\sqrt2}(|01\rangle\pm|10\rangle)$ and $\lambda_i$'s($\geq0$) are eigenvalues with $\lambda_{1,3}=\frac{1}{4}(1\pm c_1\mp c_2+c_3)$, $\lambda_{2,4}=\frac{1}{4}(1\pm c_1\pm c_2-c_3)$. We will consider the system environment interaction through operator sum representation formalism and both qubits are under similar noise. So, the dynamical evolution of the state $\rho$ under decoherence, described by a completely positive trace preserving channel $\$$, is given by,
\begin{equation}
\$(\rho)=\sum_{i,j}(E_i\otimes E_j)\rho(E_i\otimes E_j)^\dagger
\end{equation}
where $\{E_k\}$ is the set of Kraus operators associated with the decohering process of a single qubit with trace preserving condition $\sum_k{E_k^\dagger E_k}=\textit{I}.$ Here we will consider \textit{Phase Flip} (PF), \textit{Bit Flip} (BF), \textit{Bit-phase Flip} (BPF) channels as our noise models for decoherence and corresponding Kraus operators are summarized in Table(\ref{table1}).  As under every such decoherence $\rho$ preserves its Bell diagonal form, the state after evolution is given by
\begin{equation}\label{bell diagonal state after decoherence}
\rho(p)=\$(\rho)=\frac{1}{4}[I_4+\sum_{i=1}^3{c_i(p)\sigma^i\otimes \sigma^i}]=\sum_{i=1}^4{\lambda_i(p)|\phi_i\rangle\langle\phi_i|}
\end{equation}
where $\lambda_i(p)$'s are eigenvalues  of the state $\rho(p)$ with
\begin{equation}\label{eigenvalues after decoherence}
\begin{split}
\lambda_{1,3}(p)=\frac{1}{4}(1\pm c_1(p)\mp c_2(p)+c_3(p))\\ \lambda_{2,4}(p)=\frac{1}{4}(1\pm c_1(p)\pm c_2(p)-c_3(p))
\end{split}
\end{equation}
and $\rho(0)\equiv \rho$. Time dependent correlation functions $c_i(p)$'s are given in the Table(\ref{table1}) in terms of parameters $c_i(0)\equiv c_i$ of initial state and parameterized time $p=1-d(t)$ ($0 \le p \le 1$) of channel $\$$ where $d(t)$ depicts the degradation of coherence w.r.t. time $t$ and it often takes an exponential form for Markovian decoherence process.\\
\begin{table}[htp]
\begin{center}
\begin{tabular}{|c|c|c|c|c|}
\hline
Channel & Kraus operators & $c_1(p)$ & $c_2(p)$ &$c_3(p)$ \\
\hline
Phase flip &$E_1=\sqrt{1-p/2}\textit{I}$, $E_2=\sqrt{p/2}\sigma_3$ &$(1-p)^2c_1$ &$(1-p)^2c_2$ &$c_3$ \\
Bit flip &$E_1=\sqrt{1-p/2}\textit{I}$, $E_2=\sqrt{p/2}\sigma_1$ & $c_1$ &$(1-p)^2c_2$ & $(1-p)^2c_3$ \\
Bit-phase flip & $E_1=\sqrt{1-p/2}\textit{I}$, $E_2=\sqrt{p/2}\sigma_2$ &$(1-p)^2c_1$ &$c_2$ & $(1-p)^2c_3$ \\
\hline
\end{tabular}\\
\end{center}
\caption{Kraus operators $E_i$ and correlation functions $c_i(p)$ for phase flip, bit flip and bit-phase flip channels in terms of $p$ and $c_i$.}
\label{table1}
\end{table}

Since the evolutions of the state $\rho(p)$ under bit flip and bit phase flip channel are symmetric with that of phase flip channel, we will consider only phase flip channel  as noise model. Now we rename $c_1(p)$ and $c_2(p)$ by $c_+(p)$ and $c_-(p)$ in such a way that $|c_+(p)|$ and $|c_-(p)|$ are  the maximum and minimum of $\{|c_1(p)|,|c_2(p)|\}$ respectively and $c_i\equiv c_i(0)|_{i=+,-,1,2,3}$.\\\\

\noindent{\textbf{Quantum Discord under phase flip channel:}}
As the state $\rho(p)$ given in Eq.(\ref{bell diagonal state after decoherence}) is in Bell diagonal form, its total correlation $I(\rho(p))$ and classical correlation $C(\rho(p))$ are given by \cite{QD of bell state}
\begin{equation}\label{TC after decoherence}
 I(\rho(p))=2+\sum_{i=1}^4\lambda_i(p)\log_2{\lambda_i(p)}
\end{equation}
\begin{equation}\label{CC after decoherence}
C(\rho(p))=1+\sum_{i=1}^2\frac{1+(-1)^i\theta}{2}\log_2\frac{1+(-1)^i\theta}{2}
\end{equation}
respectively where $\theta \equiv \max{\{|c_1(p)|,|c_2(p)|,|c_3(p)|\}}=\max{\{|c_+(p)|,|c_3|\}}$.

The necessary and sufficient condition \cite{NASC} (in terms of correlation functions) for freezing phenomena of quantum discord for Bell diagonal state in Eq. (\ref{bell diagonal state}) under this channel  can be obtained as
\begin{equation}\label{NASC for freezing under PF}
|c_+|\ge |c_3| ,\quad c_-=-c_+c_3
\end{equation}
Now we focus on the class of initial states $\rho$ satisfying the condition in Eq. (\ref{NASC for freezing under PF}). From Eq.(\ref{eigenvalues after decoherence}) and (\ref{TC after decoherence}) it is easy to observe that total correlation of this class of states takes the form

\begin{equation}\label{TC for freezing after decoherence}
I(\rho(p))=\sum_{i=1}^2\frac{1+(-1)^ic_3}{2}\log_2[1+(-1)^ic_3]+\sum_{i=1}^2\frac{1+(-1)^i|c_+(p)|}{2}\log_2[1+(-1)^i|c_+(p)|]
\end{equation}

As under this channel $c_1(p)$ and $c_2(p)$ display the same decay rate w.r.t. $p$, $\theta=|c_+(p)|$ until a parameterized time $p_{sc}=1-\sqrt{\frac{|c_3|}{|c_+|}}$ and after that $\theta=|c_3|$. From  (\ref{CC after decoherence}) and Eq.(\ref{TC for freezing after decoherence}) it is clear that when $p \le p_{sc}$, classical correlation $C(\rho(p))$ decays monotonically and coincides with the 2nd term of total correlation $I(\rho(p))$.
\begin{figure}[htb]\label{fig:QD}
\begin{center}
\includegraphics[width=0.48\textwidth]{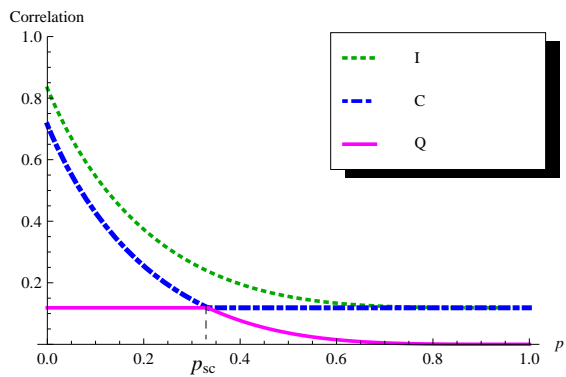}
\end{center}
\caption{\emph{Fig(1) describes the evolution of QD ($Q$) of the state with $c_1=0.9$, $c_2=-0.36$, $c_3=0.4$ under phase flip channel. Freezing of QD (magenta solid line) appears for $0\le p \le 0.333333(=p_{sc})$ with a sudden transition in its decay rate at $p=0.333333$.\\}}
\end{figure}

Therefore, for $0 \le p\le p_{sc}$ quantum discord $Q(\rho(p))$ equals to the first term in Eq(\ref{TC for freezing after decoherence}) which is constant, i.e., decay rate of quantum discord is zero for finite period of time. This behavior of QD  is known as freezing phenomena. On the other hand when $p \ge p_{sc}$, $C(\rho(p))$ in Eq(\ref{CC after decoherence}) is constant since then $\theta=|c_3|$ and hence QD $Q(\rho(p))$ decays monotonically. Such evolution of QD indicates abrupt transition in its decay rate w.r.t. parameterized time $p$ at a specific point $p=p_{sc}$. Thus QD exhibits freezing behavior for $0 \le p \le p_{sc}$ with a sudden change at $p=p_{sc}$. This dynamics is illustrated in Fig(1) where we have chosen $c_1=0.9$, $c_2=-0.36$, $c_3=0.4$. We have observed that freezing is obtained for $0 \le p \le 0.333333$ with a sudden change at $p=0.333333$.\\

\noindent{\textbf{One-norm Geometric Quantum Discord under phase flip channel:}} 
As  $\rho(p)$ given in Eq(\ref{bell diagonal state after decoherence}) is a Bell 
diagonal state, Eq.(\ref{1-GQD}) reduces to \cite{1-GQD of bell state}
\begin{equation}\label{1-GQD of bell state}
 D_G= \frac{1}{2}\times \{\text{intermediate \, value \, of}\,  \{|c_1(p)|,|c_2(p)|,|c_3|\}\}
\end{equation}
As $c_1(p)$ and $c_2(p)$ display the same decay rate w.r.t. $p$, they do not cross each other. Therefore sudden change (or double sudden change) in decay rate of $D_G$ occurs due to the allowed crossing of $|c_3|$ with either $|c_1(p)|$ or $|c_2(p)|$ (or both). Now we consider two classes of Bell diagonal states $\rho(p)$  depending upon two types of initial conditions as follows:
\textbf{Type 1:}  if $|c_+|>|c_3|>|c_-|$  and \textbf{Type 2: }   if $|c_-|>|c_3|$. \\

For type 1 states Eq.(\ref{1-GQD of bell state}) reduces to
\begin{equation}\label{1-GQD for type 1 after decoherence}
 D_G=\frac{1}{2}\begin{cases} |c_3| &\mbox{if }0\le p \le p_{sc} \\
 |c_+(p)| & \mbox{if }p_{sc}\le p \le 1. \end{cases}
\end{equation}
It is clear from Eq(\ref{1-GQD for type 1 after decoherence}) that $D_G$ exhibits freezing phenomena  for $ 0\le p \le p_{sc} $ and decays monotonically for $p_{sc}\le p\le 1$. Thus  a sudden change in its decay rate occurs at  $p=p_{sc}=1-\sqrt{\frac{|c_3|}{|c_+|}}$, caused by the only allowed crossing  $|c_3|=|c_+(p)|$.\\

For type 2 states Eq.(\ref{1-GQD of bell state}) takes the form
\begin{equation}
 D_G=\frac{1}{2}\begin{cases}|c_-(p)| &\mbox{if }0\le p \le p_{sc_1} \\
 |c_3| &\mbox{if }p_{sc_1}\le p \le p_{sc_2} \\
 |c_+(p)| & \mbox{if }p_{sc_2}\le p \le 1. \end{cases}
\end{equation}
Here the crossings $|c_3|=|c_-(p)|=(1-p)^2|c_-|$ and $|c_3|=|c_+(p)|=(1-p)^2|c_+|$ are allowed and these imply sudden transitions in decay rates of 1-GQD at two parameterized times  $p_{sc_1}=1-\sqrt{\frac{|c_3|}{|c_-|}}$ and $p_{sc_2}=1-\sqrt{\frac{|c_3|}{|c_+|}}$ which is known as double sudden change behavior.
Hence for these type of states 1-GQD exhibits freezing for $p_{sc_1}\le p \le p_{sc_2}$ with double sudden changes at $p_{sc_1}$ and $p_{sc_2}.$

\begin{figure}[htb]
\subfloat[]{\includegraphics[width=.4\textwidth,scale=0.5]{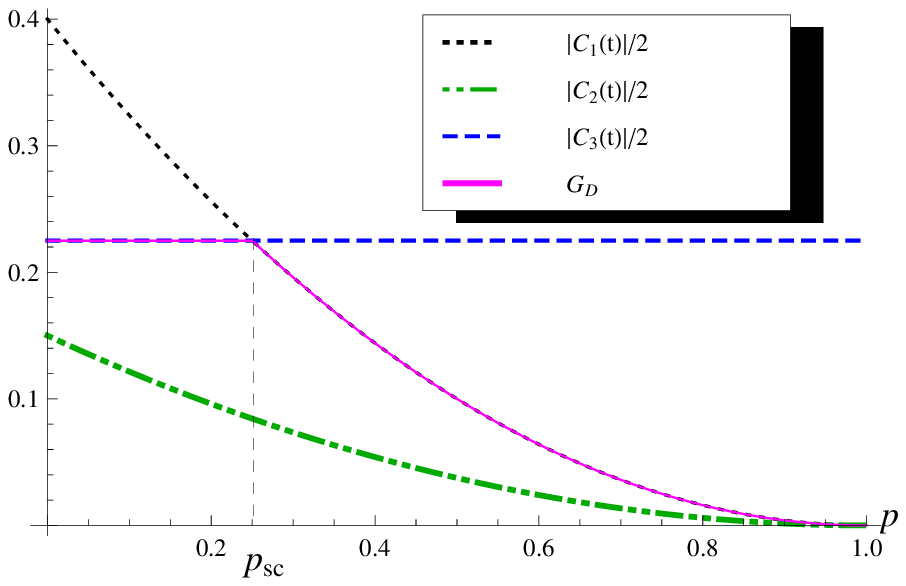} \label{fig:PF-sc}}\qquad
\subfloat[]{\includegraphics[width=.4\textwidth,scale=0.5]{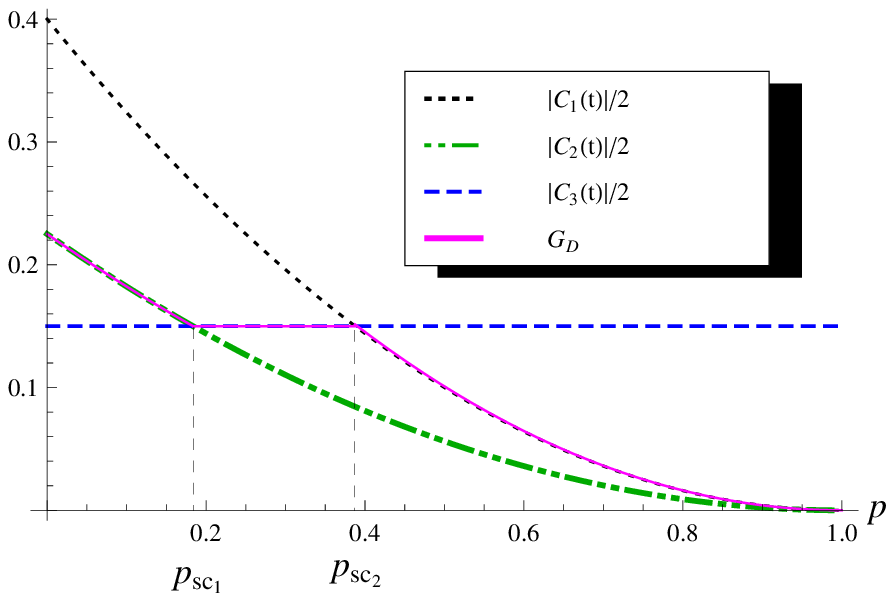}\label{fig:PF-dsc}}
\caption{\emph{ Fig(\ref{fig:PF-sc}) shows the evolution of 1-GQD($D_G$) for a type 1 state with $c_1=0.8$, $c_2=0.3$, $c_3=-0.45$. In this dynamics $D_G$(magenta solid line) becomes frozen for $0 \le p \le p_{sc}$ with single sudden change at $p=p_{sc}=0.25$ where as Fig(\ref{fig:PF-dsc}) describes the time evolution of $D_G$ for type 2  state with  $c_1=0.8$, $c_2=-0.45$, $c_3=0.3$. Under this evolution $D_G$ (magenta solid line) exhibits freezing phenomena  for  $p_{sc_1} \le p \le p_{sc_2}$ with double sudden changes at two parameterized times $p_{sc_1}=0.183503$ and $p_{sc_2}=0.387628$.} }
\label{fig:GQD under PF channel Before Filtering}
\end{figure}

In Fig(\ref{fig:GQD under PF channel Before Filtering}) we describe these dynamical evolutions of 1-GQD($D_G$) under phase flip channel for both type 1 and type 2 states by taking initial state parameters as $c_1=0.8$, $c_2=0.3$, $c_3=-0.45$ and $c_1=0.8$, $c_2=-0.45$, $c_3=0.3$ respectively. We have observed that for both the type 1 and type 2 states  QD shows freezing for finite period of time but for type 1 state it exhibits a single sudden transition at $p=0.25$ where as double sudden changes appear at $p=0.183503$ and $p=0.387628$ for type 2 state.

\section{Dynamical evolution of quantum correlations after single local filtering}
In this section, we will discuss the effect of using local filtering on the dynamical evolution of quantum correlations.
Local filtering has been seen to be capable of revealing hidden nonlocality for certain classes of states \cite{revealing entanglement1,revealing entanglement2} and useful for creation as well as purification of entanglement\cite{purification1,purification2,purification3}. Practically this map can be realized as a null-result weak measurement\cite{weak measurement}.

Let us now perform a single local filtering operation on the subsystem $A$ of state (\ref{bell diagonal state after decoherence}) and the state after filtering is given by
\begin{equation}\label{state after filtering}
\rho^k(p)=(F\otimes I)\rho(p)(F\otimes I)
\end{equation}
where filtering operator F is a non-trace preserving map and can be written as
\begin{equation}\label{filtering operation}
F=\sqrt{1-k}|0\rangle \langle0|+\sqrt{k}|1\rangle \langle1|,\qquad 0<k<1.
\end{equation}
Eigen values of the state in Eq.(\ref{state after filtering}) are $\lambda_i^k(p)$'s($\ge 0$)with
\begin{equation}\label{eigenvalues after filtering}
\begin{split}
\lambda^k_{1,3}(p)=\frac{1}{4}[(1+c_3(p))\pm\sqrt{(1-2k)^2(1-c_3(p))^2+4k(1-k)(c_1(p)-c_2(p))^2}]\\
\lambda^k_{2,4}(p)=\frac{1}{4}[(1+c_3(p))\pm\sqrt{(1-2k)^2(1-c_3(p))^2+4k(1-k)(c_1(p)-c_2(p))^2}]
\end{split}
\end{equation}
We now discuss the effects case by case.\\

\noindent{\textbf{Quantum Discord after filtering:}}
The total correlation and classical correlation of the states
$\rho^k(p)$ are given by \cite{QD of x state}
\begin{equation}\label{TC after filtering}
I(\rho^k(p))=S(\rho_A^k(p))+S(\rho_B^k(p))+\sum_{i=1}^4\lambda_i^k(p)\log_2{\lambda_i^k(p)}
\end{equation}
\begin{equation}\label{CC after filtering}
C(\rho^k(p))=S(\rho_B^k(p))+\sum_{i=1}^2\frac{1+(-1)^i\beta}{2}\log_2\frac{1+(-1)^i\beta}{2}
\end{equation}
respectively where $$\beta= \sqrt{c_3^2(1-2k)^2+4k(1-k)\theta^2},$$
$$ S(\rho_A^k(p))=-(1-k)\log_2(1-k)-k\log_2k ,$$ $$S(\rho_B^k(p))=\sum_{i=1}^2\frac{1+(-1)^i(1-2k)c_3(p)}{2}\log_2\frac{1+(-1)^i(1-2k)c_3(p)}{2}.$$
 $\lambda_i^k(p)$'s are the eigen values of state $\rho^k(p)$.
Now for this particular class of states (state $\rho$ with initial condition given in Eq(\ref{NASC for freezing under PF})) the mutual information in Eq(\ref{TC after filtering}) becomes
\begin{equation}\label{TC for freezing after filtering}
I(\rho^k(p))=S(\rho^k_A(p))+\sum_{i=1}^2\frac{1+(-1)^ic_3}{2}\log_2[1+(-1)^ic_3]+S(\rho^k_B(p))+\sum_{i=1}^2\frac{1+(-1)^i\alpha}{2}\log_2[1+(-1)^i\alpha]
\end{equation}
where $\alpha= \sqrt{(1-2k)^2+4k(1-k)c_+(p)^2}$.  Keeping in mind the values of $|c_+(p)|$ and $\theta$ it is straightforward to observe from Eq.(\ref{TC for freezing after filtering}) and Eq.(\ref{CC after filtering}) that when $p\le p_{sc}$, for each value of filtering parameter $k$  mutual information $I(\rho^k(p))$ and classical correlation $C(\rho^k(p))$ become two different monotonically decreasing function of $p$ and decreasing rate of $I$ is grater than that of $C$. Hence quantum discord is a monotonically decreasing function of $p$. On the other hand, when $p \ge p_{sc}$, $|\theta|=|c_3|$ which implies constant classical correlation $C(\rho^k(p))$, quantum discord decays monotonically. Thus a sudden transition in its decay rate appears  at $p=p_{sc}$.

Therefore local filtering affects the freezing behavior of quantum discord by removing freezing but  the point of sudden change ($p_{sc}$) remains same. In Fig(3) we describe such type of dynamical evolution of quantum discord for a state with  $c_1=0.9$, $c_2=-0.36$, $c_3=0.4$\\

\begin{figure}[htb]
\begin{center}
\includegraphics[]{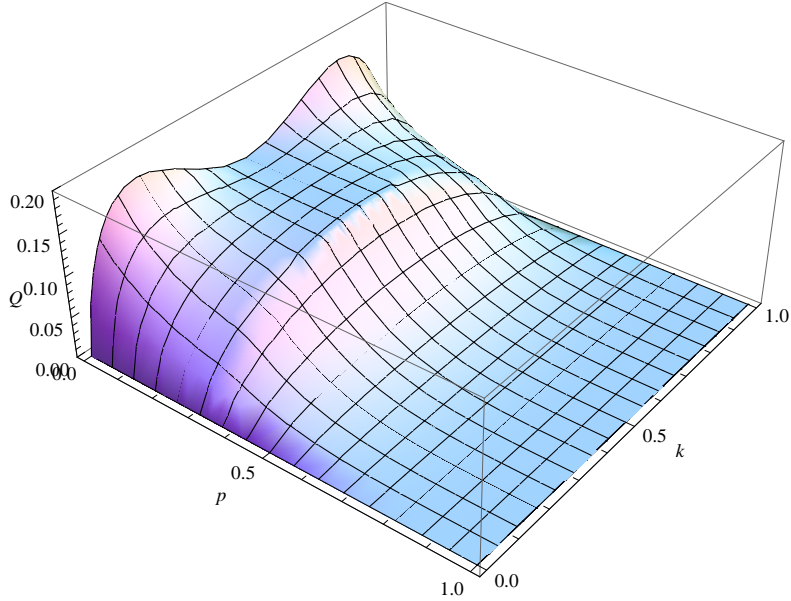}\label{fig:QD under PF-af}\\
\end{center}
\caption{\emph{Fig(3) describes the evolution of QD ($Q$) of the state $\rho$ with $c_1=0.9$, $c_2=-0.36$, $c_3=0.4$ after filtering respectively. In this dynamics freezing of QD(Q) disappears totally for all values of filtering parameter $k(\ne 0.5)$ where as before filtering for the same state QD exhibits freezing  for $0\le p \le 0.333333(=p_{sc})$ (see Fig(1)) under phase flip channel.  The sudden change in the decay rate of QD occurs at the parameterized time p=0.333333 in both before and after filtering dynamics.\\}}
\end{figure}

\noindent{\textbf{One-norm Geometric Quantum Discord after filtering:}}
As the state $\rho^k(p)$ is a X-state its 1-GQD is given by\cite{1-GQD of x state}
\begin{equation}\label{1-GQD of x state for freezing}
 D_G^k=\frac{1}{2}\sqrt{\frac{a_1 \max\{a_3,a_2\}-(a_2-q)\min\{a_3, a_1\}}{ \max\{a_3,a_2\}-\min\{a_3, a_1\}+a_1-(a_2-q)}}
\end{equation}
where $a_1=(1-q)c_+^2(p),a_2=(1-q)c_-^2(p)+q,a_3=c_3^2 $ and $q=(1-2k)^2$($0<q<1$) is the filtering parameter. The explicit expressions of $D_G^k$ depending upon parameter $q$ for type 1 and type 2 are given in Table(\ref{table2}).

\begin{table}[htp]
\begin{center}
\begin{tabular}{|p{1.1cm}|c|c||p{6.7cm}| }
\hline
\cline{1-3} State & $D_G$ & $D_G^k$,$q=(1-2k)^2, 0<k<1$ &\multicolumn{1}{c|}{$g_i(k,p),(i=1,2,3,4)$} \\
\hline
Type 1 & $ D_G=\frac{1}{2}\begin{cases} |c_3| &\mbox{if }0\le p \le p_{sc} \\
   |c_+(p)| & \mbox{if }p_{sc}\le p \le 1. \end{cases}$  &$ D_G^k=\begin{cases} g_1(k,p) &\mbox{if }0\le q \le q_1 \\
      g_2(k,p) & \mbox{if }q_1\le q \le q_2 \\
      g_3(k,p) & \mbox{if }q_2\le q \le q_3 \\
       g_4(k,p) & \mbox{if }q_3\le q < 1 \end{cases}$ &$g_1(k,p)=\frac{1}{2}\begin{cases} |c_3| &\mbox{if }0\le p \le p^k_{sc} \\
          \sqrt{1-q}|c_+(p)| & \mbox{if }p^k_{sc}\le p \le 1. \end{cases}\quad$ $g_2(k,p)=\frac{1}{2}\begin{cases} f(k,p) & \mbox{if }0\le p \le p^k_{sc_1}\\
          |c_3| & \mbox{if }p^k_{sc_1}\le p \le p^k_{sc_2} \\
                    \sqrt{1-q}|c_+(p)| & \mbox{if } p^k_{sc_2}\le p \le 1. \end{cases}$\\

 \cline{1-3}
 Type 2 &$ D_G=\frac{1}{2}\begin{cases}|c_-(p)| &\mbox{if }0\le p \le p_{sc_1} \\
    |c_3| &\mbox{if }p_{sc_1}\le p \le p_{sc_2} \\
   |c_+(p)| & \mbox{if }p_{sc_2}\le p \le 1. \end{cases}$ &$ D_G^k=\begin{cases}
        g_2(k,p) & \mbox{if }0\le q \le q_4 \\
         g_3(k,p) & \mbox{if }q_4\le q \le q_5 \\
          g_4(k,p) & \mbox{if }q_5\le q < 1 \end{cases}$ &$g_3(k,p)=\frac{1}{2}\begin{cases} f(k,p) &\mbox{if }0\le p \le p^k_{sc} \\
                    \sqrt{1-q}|c_+(p)| & \mbox{if }p^k_{sc}\le p \le 1. \end{cases}\quad$ $g_4(k,p)=\frac{1}{2}
                           \sqrt{1-q}|c_+(p)|\quad,\quad 0\le p \le 1\quad$  where  for a fixed $k$, $f(k,p)=\sqrt{\frac{a_1a_2-(a_2-q)a_3}{a_1-a_3+q}}$ is  monotonic decreasing function for $0\le p\le p^k_{sc}$ (or $ p^k_{sc_2}$). \\
\hline
\hline

 Values of $p_{sc_i}$, $\quad p^k_{sc_i}(i=$ &\multicolumn{3}{c|}{$p_{sc_1}=1-\sqrt{\frac{|c_3|}{|c_-|}}$,$\quad p_{sc_2}=1-\sqrt{\frac{|c_3|}{|c_+|}}=p_{sc}$,$\quad p^k_{sc_1}=1-[\frac{c_3^2-q}{(1-q)c_-^2}]$,$\quad p^k_{sc_2}=1-[\frac{c_3^2}{(1-q)c_+^2}]=p^k_{sc}$}  \\

$1,2)$and $ q_j$, $(j=\quad1,2,3,4)$ &\multicolumn{3}{c|}{ $q_1=\min\{\frac{c_3^2-c_-^2}{1-c_-^2},\frac{c_+^2-c_3^2}{c_+^2}\}$,$q_2  =\min\{\frac{c_3^2(c_+^2-c_-^2)}{c_+^2},\frac{c_+^2-c_3^2}{c_+^2}\}$,$q_3=\frac{c_+^2-c_3^2}{c_+^2}$,$q_4=\frac{c_3^2(c_+^2-c_-^2)}{c_+^2}$ and $q_5=\frac{c_+^2-c_3^2}{c_+^2}$}
    \\

  \hline
  \end{tabular}\\
  \end{center}
  \caption{Explicit expressions of 1-GQD before and after filtering ($D_G$ and $D_G^k$ respectively) for type 1 and type 2 states.}
  \label{table2}
  \end{table}

\begin{figure}[htb]
\subfloat[]{\includegraphics[trim = 0mm 0mm 0mm 0mm,clip,scale=0.6]{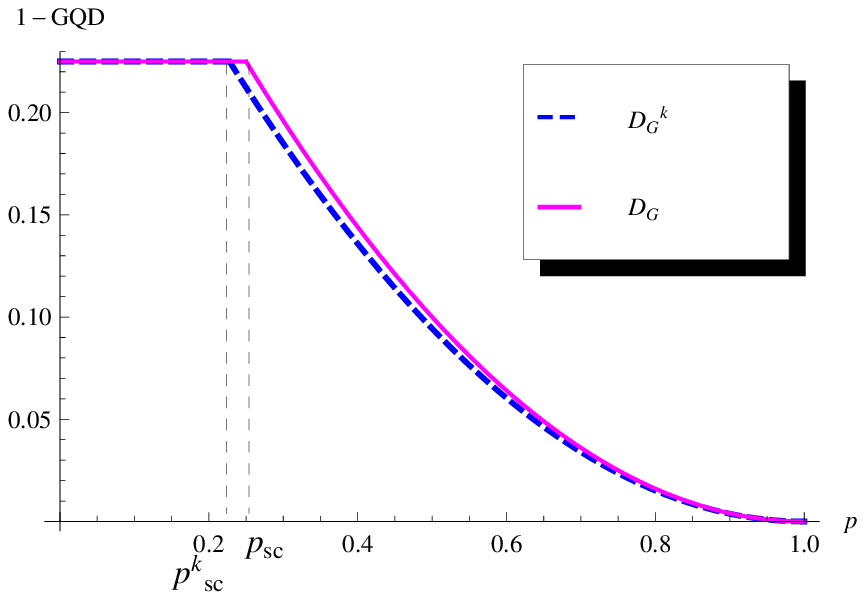} \label{fig:PF-sc q=0.11}}\quad
\subfloat[]{\includegraphics[trim = 0mm 0mm 0mm 0mm,clip,scale=0.6]{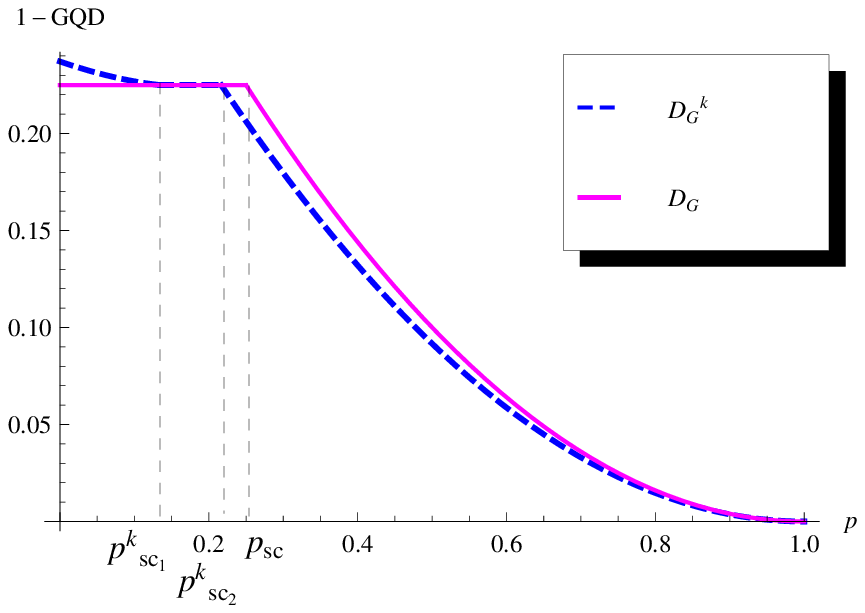}\label{fig:PF-sc q=0.16}}\\
\subfloat[]{\includegraphics[trim = 0mm 0mm 0mm 0mm,clip,scale=0.6]{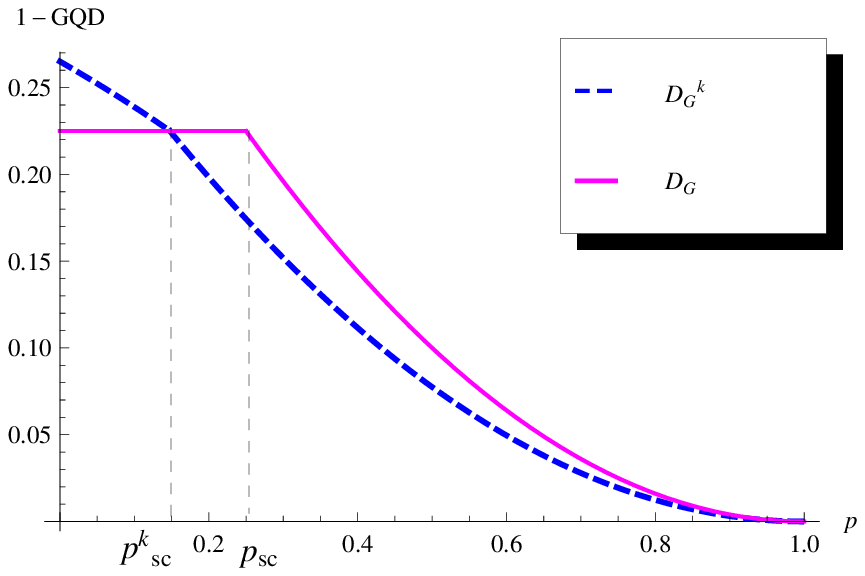} \label{fig:PF-sc q=0.4}}\qquad
\subfloat[]{\includegraphics[trim = 0mm 0mm 0mm 0mm,clip,scale=0.6]{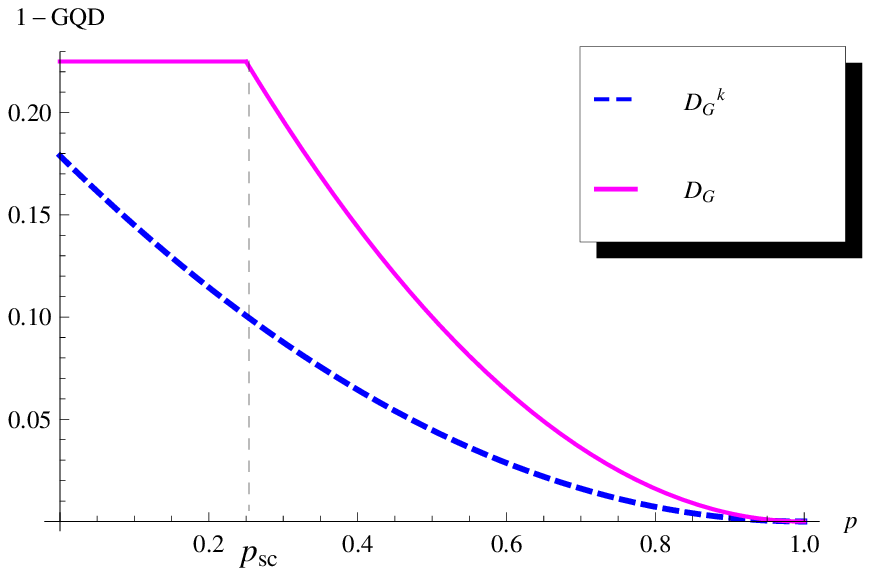} \label{fig:PF-sc q=0.8}}\\
\subfloat[]{\includegraphics[trim = 0mm 0mm 0mm 0mm,clip,scale=0.6]{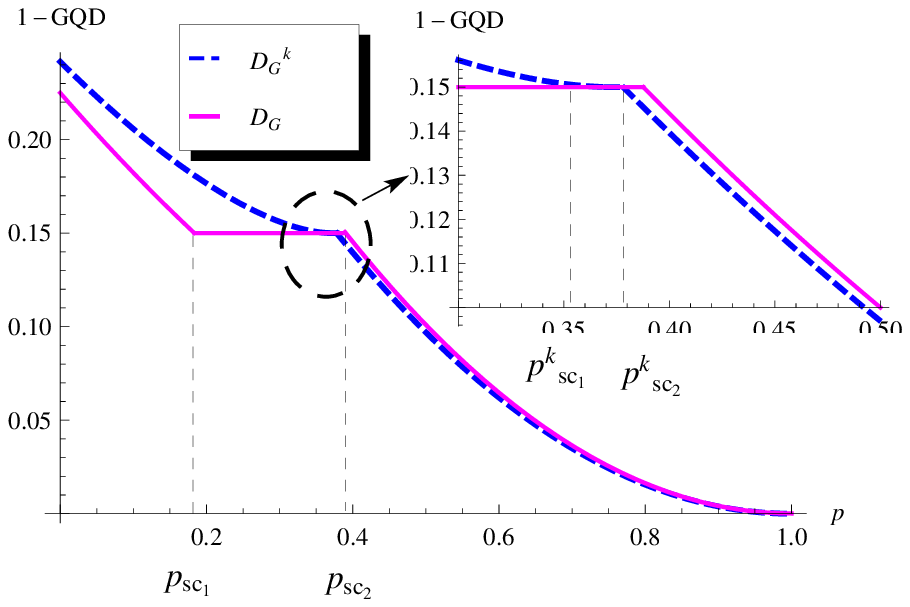} \label{fig:PF-dsc q=0.06}}
\subfloat[]{\includegraphics[trim = 0mm 0mm 0mm 0mm,clip,scale=0.6]{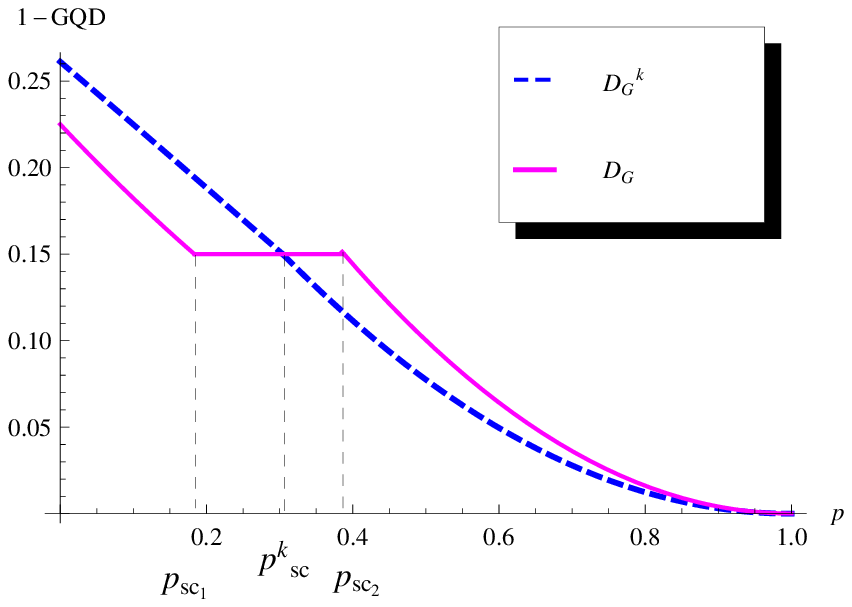} \label{fig:PF-dsc q=0.4}}
\subfloat[]{\includegraphics[trim = 0mm 0mm 0mm 0mm,clip,scale=0.6]{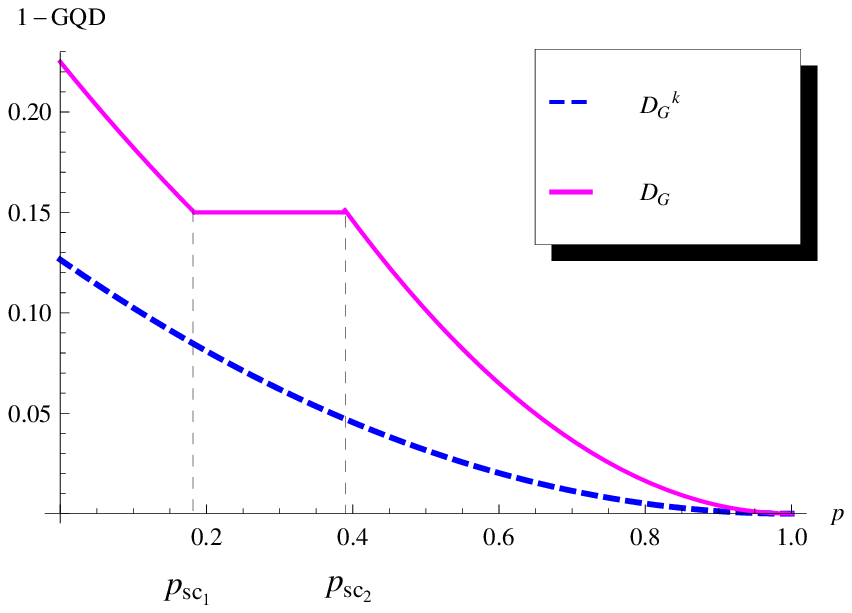} \label{fig:PF-dsc q=0.9}}
\caption{\emph{Here we describes the evolution of $D_G$(1-GQD before filtering)(blue dotted line) and $D_G^k$(1-GQD after filtering)(magenta solid line) for both the type 1 and type 2 states. Fig(\ref{fig:PF-sc q=0.11}),Fig(\ref{fig:PF-sc q=0.16}),Fig(\ref{fig:PF-sc q=0.4}) and Fig(\ref{fig:PF-sc q=0.8}) are for the examples of type 1 states with $c_1=0.8,c_2=0.3,c_3=-0.45$. For this state $D_G$ remains frozen for $0\le p\le 0.25(=p_{sc}).$ After filtering for $q=0.11$ (Fig(\ref{fig:PF-sc q=0.11})) freezing of $D_G^k$ appears for $0\le p\le 0.227829(=p^k_{sc})$ i.e., duration of freezing of $D_G^k$ is less than that of $D_G$. For $q=0.16$ (Fig(\ref{fig:PF-sc q=0.16})) this freezing  appears for $p^k_{sc_1}\le p \le p^k_{sc_2}$ with double sudden changes  at $p^k_{sc_1}=0.134102$ and $p^k_{sc_2}=0.216586$. When $q=0.4$ (Fig(\ref{fig:PF-sc q=0.4})) freezing of 1-GQD totally disappears but a single sudden change is seen at $p^k_{sc}=0.147835$ and when $q=0.8$ (Fig(\ref{fig:PF-sc q=0.8}))$D_G^k$ becomes monotonic without any sudden change. Fig(\ref{fig:PF-dsc q=0.06}),Fig(\ref{fig:PF-dsc q=0.4}),Fig(\ref{fig:PF-dsc q=0.9}) shows the evolution of 1-GQD for type 2 states with $c_1=0.8,c_2=-0.45,c_3=0.3$. For this state  $D_G$ exhibits freezing behavior for finite time interval $[p_{sc_1},p_{sc_2}]$ with double sudden changes at $p_{sc_1}=0.183503$ and $p_{sc_2}=0.387628$. After filtering for $q=0.06$ (Fig(\ref{fig:PF-dsc q=0.06})) $D_G^k$ remains constant or frozen for $p^k_{sc_1}\le p \le p^k_{sc_2}$ with $p^k_{sc_1}=0.369925$ and $p^k_{sc_2}=0.378081$, i.e., time of freezing becomes less than that of before filtering . For $q=0.4$ (Fig(\ref{fig:PF-dsc q=0.4})) a sudden change occurs at $p^k_{sc}=0.304211$ but freezing of $D_G^k$ is removed and for $q=0.9$ (Fig(\ref{fig:PF-dsc q=0.9})) sudden change also disappears.}}
\label{fig:GQD under PF channel}
\end{figure}

For type 1 state 1-GQD($D_G^k$) in Eq(\ref{1-GQD of x state for freezing}) is any one of $g_i(p)$'s($i=1,2,3,4$) given in Table(\ref{table2}) depending upon values of $q$ and its decay rate exhibits any one of the following dynamics:
\begin{enumerate}
\item If $ 0<q\le q_1$(fig(\ref{fig:PF-sc q=0.11})), $D_G^k$ shows similar type of evolution as $D_G$, i.e., it is constant for a finite interval $[0,p^k_{sc}]$ of parameterized time $p$ and then decays monotonically. As $p_{sc}\le p^k_{sc}$, duration of freezing of $D_G^k$ is less than that of $D_G$.
\item If $ q_1<q\le q_2$ (fig(\ref{fig:PF-sc q=0.16})), $D_G^k$ decays monotonically until a parameterized time $p^k_{sc}$. After that it remains constant for the time interval  $p^k_{sc_1}\le p\le p^k_{sc_2}$ and then decays monotonically again. This implies two times abrupt transition in its decay rate where as, before filtering there was a single sudden change. Since $p^k_{sc_1}\le p^k_{sc_2}\le p_{sc}$, the duration of freezing ($p^k_{sc_2}-p^k_{sc_1}$) reduces more as compared with that of the previous case  $ 0<q\le q_1$.
\item If $q_2\le q \le q_3$ (fig(\ref{fig:PF-sc q=0.4})), only a single sudden transition is seen at $p=p^k_{sc}$ in its decay rate and freezing vanishes.
\item If $q_3< q <1 $ (fig(\ref{fig:PF-sc q=0.8})), $D_G^k$ decays monotonically for the whole range of $p$ without any freezing or any sudden change.
\end{enumerate}

For type 2 states depending upon the choice of filtering parameter $q$ its evolution can be any of the following types:
\begin{enumerate}
\item If $ 0<q\le q_4$(fig(\ref{fig:PF-dsc q=0.06})), then nature of decay rate of 1-GQD ($D_G^k$) is same as before filtering. $D_G^k$ becomes freezed for $p^k_{sc_1}\le p\le p^k_{sc_2}$ and sudden changes appears at the parameterized times $p^k_{sc_1}$ and $p^k_{sc_2}$. But duration of freezing of $D_G^k$ reduces as compared with that of $D_G$ because $p_{sc_1}\le p^k_{sc_1} \le p^k_{sc_2}\le p_{sc_2} $.
\item If $q_4\le q <q_5 $(fig(\ref{fig:PF-dsc q=0.4})), then freezing phenomena disappears and just a single sudden change in its decay rate is seen at $p=p^k_{sc}$.
\item If $q_5\le q <1 $(fig(\ref{fig:PF-dsc q=0.9})), then freezing, sudden change both disappear and $D_G^k$ decays monotonically.
\end{enumerate}

 From the above cases of both type 1 and type 2 states,  it is clear that for any values of filtering parameter $q$, duration of freezing reduces and this reduction increases with the increment of $q$. If we choose $q$ in such a way that $q \ge q_4$ or $q_5$, freezing and sudden change both disappear.\\

\textit{\textbf{ Remarks on Quantum Discord and 1-GQD Under Bit flip(or Bit-phase flip) channel :}} Before and after filtering, the dynamical evolution of quantum discord and 1-GQD  under BF (or BPF) channel  is symmetric with that of phase flip channel just one has to exchange $c_3$ and $c_1$ (or $c_2$). \\\\

 \
\section{CONCLUSION}
In conclusion, we have analyzed in detail, the effect of single local filtering operation on  dynamical evolution of  Quantum Discord and One Norm Geometric Quantum Discord for the Bell diagonal state under different Markovian noise such as phase flip, bit flip, bi-phase flip. During evolution under these channels few crucial features like freezing, sudden change, double sudden change behaviors of these quantum correlations are seen in its decay rate. We have shown that single local filtering is able to remove this freezing and this disappearance of freezing totally depends on the value of filtering parameter $q(0<q<1)$. In case of Quantum discord, any amount of filtering helps to remove its freezing but the sudden change in its decay rate occurs at the same parameterized time as before filtering. On the other hand in case of 1-GQD,  any amount of filtering reduces the duration of freezing. We observed that when amount of filtering increases, duration of freezing decreases and thus the points of abrupt transition in the decay rate of quantum correlation changes. We have found a range of filtering parameter, for which  1-GQD decays monotonically  without both freezing and any sudden change in its decay rate. We hope our results will help further to understand basic nature of quantum correlations beyond entanglement.

{\bf Acknowledgement.} The authors S. Karmakar, A. Sen acknowledges the financial support from University Grants Commission, New Delhi, India. Conflict of Interest: The authors declare that they have no conflict of interest.
\\

\end{document}